# A System-Level Voltage/Frequency Scaling Characterization Framework for Multicore CPUs

George Papadimitriou, Manolis Kaliorakis, Athanasios Chatzidimitriou, Dimitris Gizopoulos
University of Athens, Greece
{georgepap, manoliskal, achatz, dgizop}@di.uoa.gr

Greg Favor, Kumar Sankaran
Applied Micro
Santa Clara, CA, USA
{gfavor, ksankaran}@apm.com

Shidhartha Das
ARM
Cambridge, UK
Shidhartha.Das@arm.com

***Abstract*—Supply voltage scaling is one of the most effective techniques to reduce the power consumption of microprocessors. However, technology limitations such as aging and process variability enforce microprocessor designers to apply pessimistic voltage guardbands to guarantee correct operation in the field for any foreseeable workload. This worst-case design practice makes energy efficiency hard to scale with technology evolution. Improving energy-efficiency requires the identification of the chip design margins through time-consuming and comprehensive characterization of its operational limits. Such a characterization of state-of-the-art multi-core CPUs fabricated in aggressive technologies is a multi-parameter process, which requires statistically significant information.**

**In this paper, we present an automated framework to support system-level voltage and frequency scaling characterization of Applied Micro's state-of-the-art ARMv8-based multicore CPUs used in the X-Gene 2 micro-server family. The fully automated framework can provide fine-grained information of the system's state by monitoring any abnormal behavior that may occur during reduced supply voltage conditions. We also propose a new metric to quantify the behavior of a microprocessor when it operates beyond nominal conditions. Our experimental results demonstrate potential uses of the characterization framework to identify the limits of operation for improved energy efficiency.**

## I. Introduction

Technology evolution with shrinking transistors increases chip density throughout the entire computing spectrum, from handheld devices employed at the edge of the Internet-of-Things (IoT) to servers, datacenters and supercomputers. The gains are indisputable in terms of performance, but come at the expense of increased power consumption, which elevates energy efficiency as a primary objective of computing systems design. One of the most effective techniques to improve energy efficiency is to reduce the chip supply voltage.

During chip fabrication, process variations can affect transistor dimensions (length, width, oxide thickness etc.) [1] which have direct impact on the threshold voltage of a MOS device [2]. As technology scales, the percentage of these variations compared to the overall transistor size increases and raises major concerns for designers, who intend to increase the energy efficiency. This variation is classified as static variation and remains constant after fabrication. Additional to that, transistor aging and dynamic variation in supply voltage and temperature, caused by different workload interactions, is also of primary importance. Both static and dynamic variations lead microprocessor architects to apply aggressive guardbands (operating voltage and frequency settings) in order to avoid timing failures and guarantee correct operation, even in the worst-case conditions excited by unknown workloads [3] [4]. However, these guardbands impede the low power consumption and the high performance, which can be derived by reducing the supply voltage and increasing the operation frequency, respectively.

To bridge the gap between energy efficiency and performance improvements, several hardware and software techniques have been proposed, such as Dynamic Voltage and Frequency Scaling (DVFS) [5]. The premise of DVFS is that a microprocessor's workloads as well as the cores' activity vary, so when one or more cores have less or no work to perform, the frequency, and thus, the voltage can be slowed down without affecting performance adversely. However, to further reduce the power consumption by keeping the frequency high when it is necessary, recent studies aim to uncover the conservative operational limits, by performing an extensive system-level voltage scaling characterization of commercial microprocessors' operation beyond nominal conditions [6] [7] [8] [9] [10]. These studies leverage the Reliability, Accessibility, and Serviceability (RAS) features, provided by the hardware (such as ECC), in order to expose reduced but safe operating margins.

A major challenge, however, in voltage scaling characterization at the system level is the time-consuming large population of experiments due to: (i) different voltage and frequency levels, (ii) different characterization setups (e.g. for a multicore chip both the cases of running a benchmark in each individual core and simultaneously in all cores should be examined), and (iii) diverse-behavior workloads. In addition, due to the non-deterministic behavior of the experiments, caused by different microarchitectural events that occur in a system-level characterization and to ensure the statistical significance of the observations, the same experiments should be repeated multiple times at the same voltage level, which further increases the characterization time. Moreover, when the system operates in voltage levels that are significantly lower than its nominal value, system crashes are frequent and unavoidable and the recovery from these cases constitutes a significant portion of the overall experiment time. For these reasons, manually-controlled voltage scaling characterization is infeasible; a generic and automated experimental framework that can be easily replicated in different machines is required. Furthermore, such a framework has to ensure the credibility of the delivered results because when a system operates beyond nominal conditions it can fall in unstable states.

In this paper, we present a versatile framework to study the behavior of multicore CPUs, when they operate under scaled voltage and frequency conditions. We build a voltage/frequency (V/F) scaling characterization framework on Applied Micro's (APM) X-Gene 2 micro-server family, fabricated in 28nm process technology that consists of eight ARMv8-compliant cores. The proposed infrastructure is fully aligned with all the aforementioned requirements and aims to investigate the limits of a state-of-the-art microprocessor architecture beyond the nominal conditions: it is fast, reliable and easily extensible and replicable. The characterization framework records several different types of deviations from the normal execution as proxies for the effect of voltage and frequency scaling.

We also propose a metric called *Severity Function*, to both quantify the severity and illustrate the scaling of abnormal behaviors due to voltage reduction. The metric's contribution is twofold: (1) to aggregate the results produced by multiple runs, and (2) to quantify a microprocessor's ability to operate beyond nominal conditions. To the best of our knowledge, this is the first study that presents in detail an automated infrastructure for the characterization of ARM-based systems beyond nominal conditions.

The rest of the paper is organized as follows. Section II describes all the related works, while Section III describes all the details of the X-Gene 2 and our proposed framework. Finally, Section IV presents our experimental results and Section V summarizes all our conclusions.

## II. Related Work

During the last years, the goal for improving microprocessors' energy efficiency, while reducing their power supply voltage is a main concern of many scientific studies that investigate the chips' operation limits in nominal and off-nominal conditions. For example, in [11] [12] and [13] the authors propose methods to maximize voltage droops in single core and multicore chips in order to investigate their worst case behavior due to the generated voltage noise effects.

In order to eliminate the effects of voltage noise, studies such as [14] and [15] focus on the prediction of critical parts of benchmarks, in which large voltage noise glitches are likely to occur, leading to system malfunctions. In the same context, several studies either in the hardware or in the software level were presented to mitigate the effects of voltage noise [4] [16] [17] [18] [19] or to recover from them after their occurrence [20].

Apart from these studies that are mainly concentrated on the core and the voltage droops, [7] and [8] focus on the observation of the errors manifested on caches of a commercial Intel Itanium processor during the execution of benchmarks in voltage conditions in off-nominal values. The authors in these studies observe how the ECC rates increase along with the supply voltage reduction. Moreover, the authors in [21] [22] and [23] propose several microarchitectural approaches to ensure the correct operation of caches in ultra-low voltage conditions.

Finally, the characterization studies of commercial chips in off-nominal voltage conditions are limited [6] [7] [8] [9] [10], strengthening the purpose of the existence of our proposed framework that targets the commercial APM X-Gene 2 micro-server (fabricated in 28nm process technology).

## III. System Architecture

The APM X-Gene 2 micro-server consists of eight 64-bit ARMv8 cores. The X-Gene 2 architecture offers high-end processing performance and capabilities. For example, the X-Gene 2 subsystem features the Power Management processor (PMpro) and Scalable Lightweight Intelligent Management processor (SLIMpro) to enable breakthrough flexibility in power management, resiliency, and end-to-end security for a wide range of applications. The PMpro, a 32-bit dedicated processor provides advanced power management capabilities such as multiple power planes and clock gating, thermal protection circuits, Advanced Configuration Power Interface (ACPI) power management states and external power throttling support. The SLIMpro, a 32-bit dedicated processor monitors system sensors, configure system attributes (e.g. regulate supply voltage, change DRAM refresh rate etc.) and access all error reporting infrastructure, using an integrated I2C controller as the instrumentation interface between the X-Gene 2 Cores and this dedicated processor. SLIMpro can be accessed by the system's running Linux Kernel.

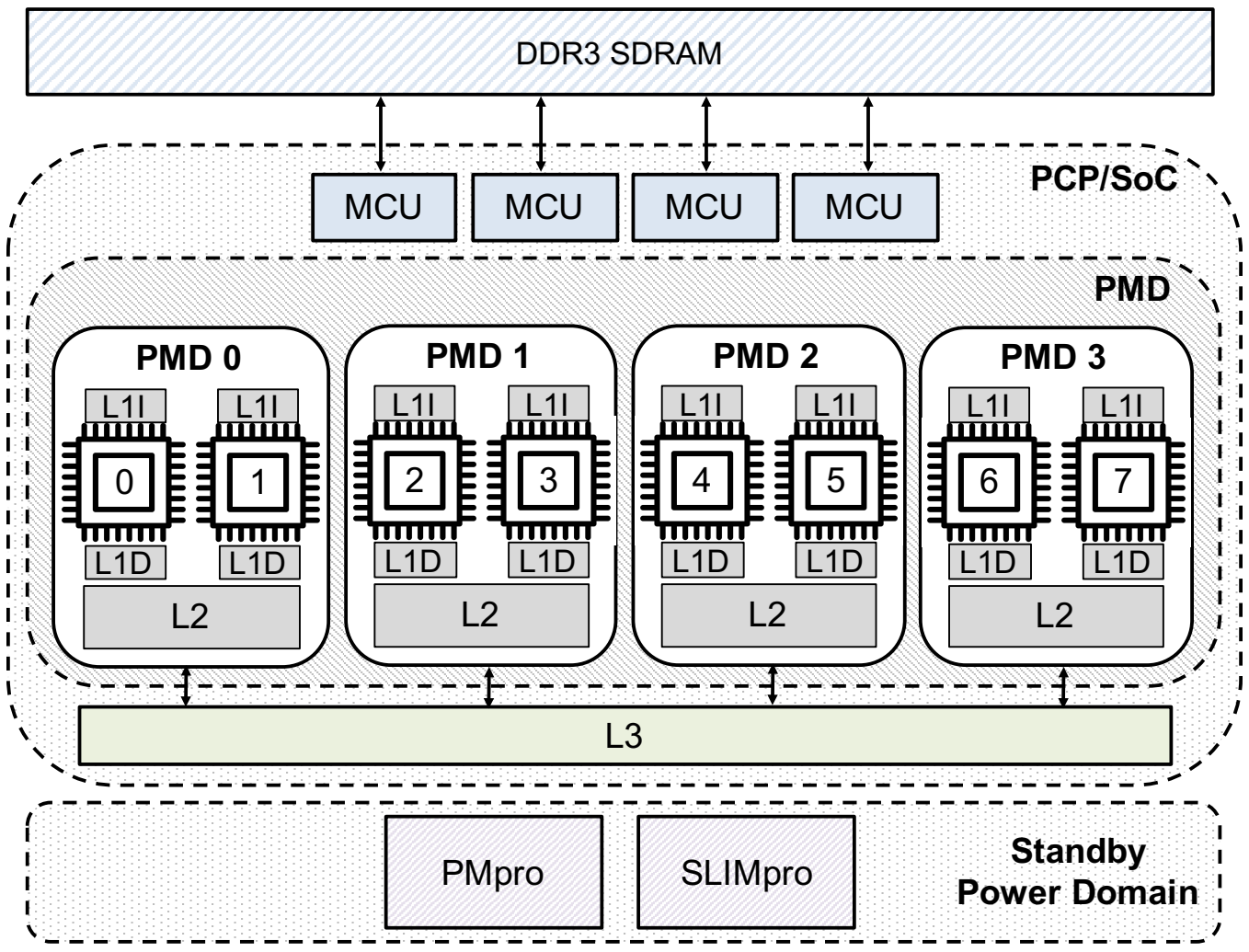

Fig. 1. X-Gene 2 micro-server power domains block diagram. The outlines with dashed lines present the independent power domains of the chip.

X-Gene 2 has three independently regulated power domains (as shown in Fig. 1 above):

1) **PMD (Processor Module)**: Each PMD contains two ARMv8 cores. Each of the two cores has separate instruction and data caches, while they share a unified L2 cache. The operating voltage of all four PMDs together can change with a granularity of 5mV beginning from 980mV. While PMDs operate at the same voltage, each PMD can operate in a different frequency. The frequency can range from 300MHz up to 2.4GHz at 300MHz steps.

2) **PCP (Processor Complex)/SoC**: It contains the L3 cache, the DRAM controllers, the central switch and the I/O bridge. The PMDs do not belong to the PCP/SoC power domain. The voltage of the PCP/SoC domain can be independently scaled downwards with a granularity of 5mV beginning from 950mV.
3) **Standby Power Domain**: This includes the SLIMpro and PMpro microcontrollers and interfaces for I2C buses.

TABLE I summarizes the most important architectural and microarchitectural parameters of the APM X-Gene 2 micro-server that is used in our study.

TABLE I: BASIC CHARACTERISTICS OF X-GENE 2.

| Parameter | Configuration |
|---|---|
| ISA | ARMv8 (AArch64, AArch32, Thumb) |
| Pipeline | 64-bit OoO (4-issue) |
| CPU | 8 Cores, 2.4GHz |
| L1 Instr. Cache | 32KB per core (Parity Protected) |
| L1 Data Cache | 32KB per core (Parity Protected) |
| L2 cache | 256KB per PMD (ECC Protected) |
| L3 cache | 8MB (ECC Protected) |

## IV. FRAMEWORK OVERVIEW

The primary goals of the proposed framework are: (1) to identify the target system's limits when it operates at scaled voltage and frequency conditions, and (2) to record/log the effects of a program's execution under these conditions. The framework provides the following features; it:

- compares the outcome of the program with the correct output of the program when the system operates in nominal conditions to record Silent Data Corruptions (SDCs),
- monitors the exposed corrected and uncorrected errors from the hardware platform's error reporting mechanisms
- recognizes when the system is unresponsive to restore it automatically,
- monitors system failures (crash reports, kernel hangs, etc.),
- determines the safe, unsafe and non-operating voltage regions for each application for all frequencies, and
- performs massive repeated executions of the same configuration.

The automated framework (outlined in Fig. 2) is easily configurable by the user, and can be embedded to any Linux-based system, with similar voltage and frequency regulation capabilities. As shown in Fig. 2, the proposed framework consists of three phases (Initialization, Execution, Parsing).

To completely automate the characterization process, and due to the frequent and unavoidable system crashes that occur when the system operates in reduced voltage levels, we set up a Raspberry Pi board connected externally to the X-Gene 2 board which behaves as a watchdog. The Raspberry is physically connected to both the Serial Port and the Power and Reset buttons of the system board to enable physical access to the system.

We discuss the several challenges that were taken into consideration for a solid development of such a framework.

**Safe Data Collection**. Given that a system operating beyond nominal conditions often has unexpected behaviors (e.g. file system driver failures), there is the need to correctly identify and store all the essential information in log files (to be subsequently parsed and analyzed). The automated framework was developed in such a way to collect and store safely all the necessary information about the experiments.

**Failure Recognition**. Another challenge is to recognize and distinguish the system and program crashes or hangs. This is a very important feature for the Parsing Phase to easily identify and classify the final results, with the most possible distinct information concerning the characterization.

**Reliable Cores Setup**. Another major challenge we also face is that the characterization of a system is performed primarily by using properly chosen programs in order to provide diverse behaviors and expose all the potential deviations from nominal conditions. It is thus important to run the selected benchmarks in *reliable cores setup*. This means that the cores, where the benchmark runs, must be isolated and unaffected from the other active processes of the kernel in order to capture only the effects of the desired benchmark.

**Iterative Execution**. The non-deterministic behavior of the characterization results due to several microarchitectural features makes necessary to repeat the experiments multiple times with the same configuration to eliminate the probability of misleading results.

In the following subsections, we analyze each of these functionalities grouped in the 3 distinct phases of the framework's execution (Initialization, Execution, Parsing), and describe their detailed implementation and how these challenges were overcome.

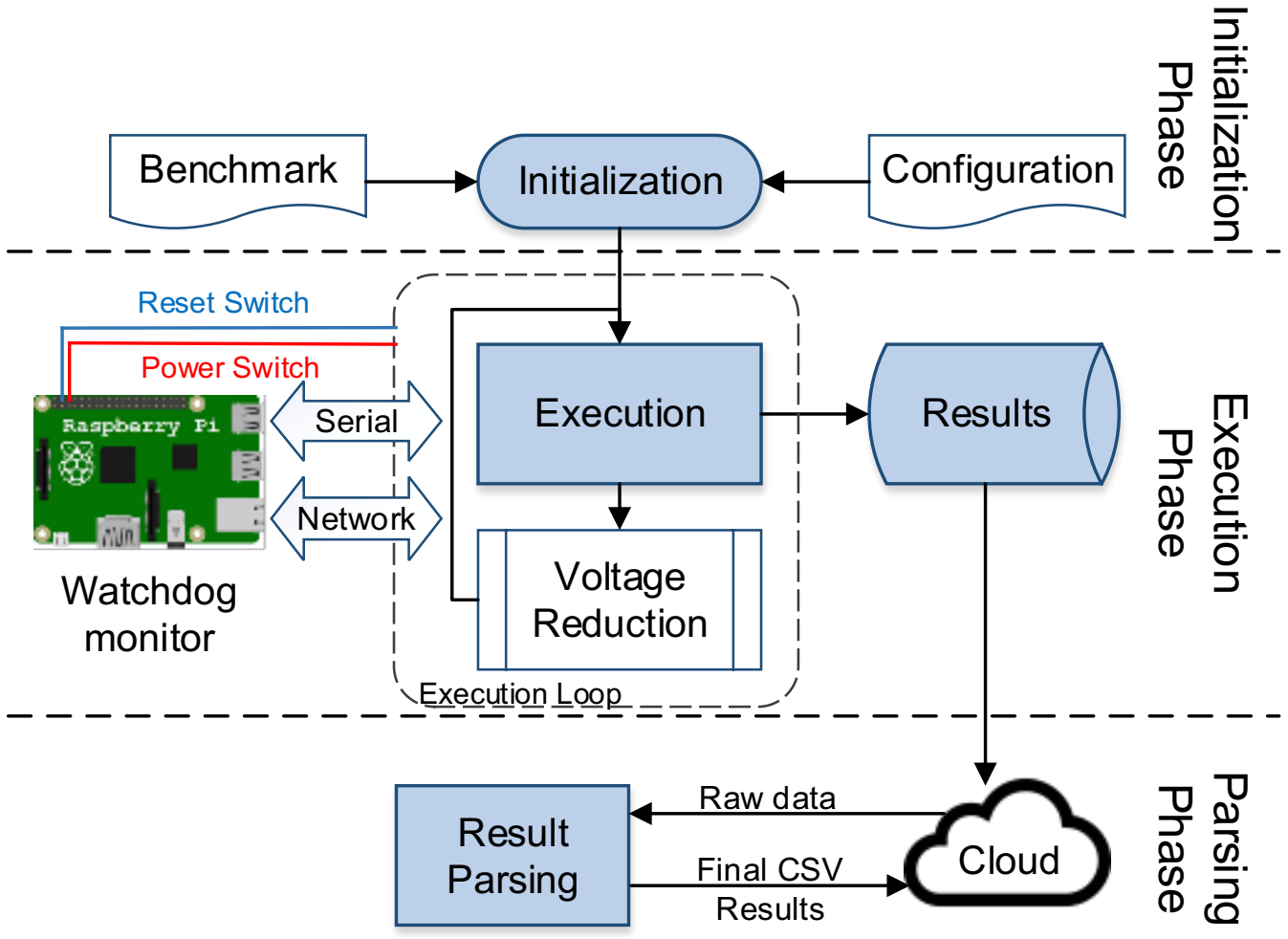

Fig. 2. Framework Layout.

### A. *Initialization Phase*

During the *initialization phase*, a user can declare a benchmark list with any input dataset to run in any desirable characterization setup. The characterization setup includes the voltage and frequency (V/F) values under which the experiment will take place and the cores where the benchmark will be run; this can be an individual core, a pair of cores in the same PMD, or all of the available eight cores in the microprocessor. The characterization setup depends on the power domains supported by the chip, but our framework is easily extensible to support the power domain features of different CPU chips.

This phase is in charge of setting the voltage and frequency ranges, the initial voltage and frequency values, with which the characterization begins, and to prepare the benchmarks: their required files, inputs, outputs, as well as the directory tree where the necessary logs will be stored. This phase is performed at the beginning of the characterization and each time the system is restored by the Raspberry (for example, after a system crash) in order to proceed to the next run until the entire Execution Phase finishes. Each time the system is restored, this phase restores the initial user's desired setup and recognizes where and when the characterization has been previously stopped. This step is essential for the characterization to proceed sequentially according to user's choice, and to complete the whole Execution Phase.

This phase is also responsible to overcome the challenge of *reliable cores setup* that is responsible to ensure the correctness and integrity of our results. The benchmark must run in an "as bare as possible" system without the interference of any other running process. Therefore, reliable cores setup is twofold: first, it recognizes these cores or group of cores that are not currently under characterization, and migrates all currently running processes (except for the benchmark) to a completely different core. The migration of system processes is required to isolate the execution of the desired benchmark from all other active processes. Second, given that all the PMDs in the studied system are in the same power domain, they always have the same voltage value (in case this does not hold in a different microarchitecture the proposed framework can be adapted). This means that even though there are several processes run on different cores (not in the core(s) under characterization), they have the same probability to affect an unreliable operation while reducing the voltage.

On the other hand, each individual PMD can have different frequency, so we leverage the combination of V/F states in order to set the core under characterization to the desired frequency, and all other cores to the minimum available frequency in order to ensure that an unreliable operation is due to the benchmark's execution only. When for example the characterization takes place in the PMD0 (meaning that the benchmark runs in PMD0; cores 0 and 1), the PMD0 is set to the pre-defined by the user frequency, and all the other PMDs are set to the minimum available frequency (300MHz in our case). Thus, all the running processes, except for the benchmark, are executed to the reliable cores setup.

In our setup, we also use a stripped/lightweight Linux Kernel to diminish the unnecessary kernel daemons that the majority of well-known Linux Distributions provide. Thus, the system's running processes and the common power domain of all PMDs, neither affect the benchmarks execution nor can contribute to a system's failure or error event.

### B. *Execution Phase*

After the characterization setup is defined, the automated *Execution Phase* begins. The Execution Phase consists of multiple runs of the same benchmark, each one representing the execution of the benchmark with a pre-defined characterization setup. The set of all the characterization runs running the same benchmark with different characterization setups represents a *campaign*. After the initialization phase, the framework enters the Execution Phase, in which all runs take place. The runs are executed according to user's configuration, while the framework reduces the voltage with a step defined by the user in the initialization phase. For each run, the framework collects and stores the necessary logs at a safe place externally to the system under characterization, which will be then used by the parsing phase.

The logged information includes: the output of the benchmark at each execution, the corrected and uncorrected errors (if any) collected by the Linux EDAC Driver [24], as well as the errors' localization (L1 or L2 cache, DRAM, etc.), and several failures, such as benchmark crash, kernel hangs, and system unresponsiveness. The framework can distinguish these types of failures and keep logging about them to be parsed later by the parsing phase. Benchmark crashes can be distinguished by monitoring the benchmark's exit status. On the other hand, to identify the kernel hangs and system unresponsiveness, during this phase the framework notifies the Raspberry when the execution is about to start and also when the execution finishes.

In the meantime, the Raspberry starts pinging the system to check its responsiveness. If the Raspberry does not receive a completion notification (hang) in the given time (we defined as timeout condition a 2 times the normal execution time of the benchmark) or the X-Gene 2 turns completely unresponsive (ping is not responding), the Raspberry sends a signal to the Power Off button on the board, and the system resets. After that, the Raspberry is also responsible to check when the system is up again, and sends a signal to restart the experiments. These decisions contribute to the *Failure Recognition* challenge.

During the experiments, some Linux tasks or the kernel may hang. To identify these cases, we use an inherent feature of the Linux kernel to periodically detect these tasks by enabling the flag "*hung_task_panic*" [24]. Therefore, if the kernel itself recognizes a process hang, it will immediately reset the system, so there is no need for the Raspberry to wait until the timeout. In this way, we also contribute to the *Failure Recognition* challenge and accelerate the reset procedure and the entire characterization.

Note that, in order to isolate the framework's execution from the core(s) under characterization, the operations of the framework are also performed in *Reliable Cores Setup*. However, when there are operations of the framework, such as the organization of log files during the benchmark's execution that are an integral part of the framework, and thus, they must run in the core(s) under characterization, these operations are performed after the benchmark's execution in the nominal conditions. This is the way to ensure that any logging information will be stored correctly and no information will be lost or changed due to the unstable system conditions, and thus, to overcome the *Safe Data Collection* challenge.

### C. Parsing Phase

In the last step of our framework, all the log files that are stored during the Execution Phase are parsed in order to provide a fine-grained classification of the effects observed for each characterization run. Note that, each run is correlated to a specific benchmark and characterization setup. The categories that are used for our classification are summarized in TABLE II, but the parser can be easily extended according to the user's needs. For instance, the parser can also report the exact location that the correctable errors occurred (e.g. the cache level, the memory, etc.) using the logging information provided by the Execution Phase.

TABLE II: EXPERIMENTAL EFFECT CATEGORIZATION.

| Effect | Description |
|---|---|
| **NO** (Normal Operation) | The benchmark was successfully completed without any indications of failure. |
| **SDC** (Silent Data Corruption) | The benchmark was successfully completed, but a mismatch between the program output and the correct output was observed. |
| **CE** (Corrected Error) | Errors were detected and corrected by the hardware. |
| **UE** (Uncorrected Error) | Errors were detected, but not corrected by the hardware. |
| **AC** (Application Crash) | The application process was not terminated normally (the exit value of the process was different than zero). |
| **SC** (System Crash) | The system was unresponsive; meaning that the X-Gene 2 is not responding to pings or the timeout limit was reached. |

Note that each characterization run can manifest multiple effects. For instance, in a run both SDC and CE can be observed; thus, both of them should be reported by the parser for this run. Furthermore, the parser can report all the information collected during multiple campaigns of the same benchmark. The characterization runs with the same configuration setup of different campaigns may also have different effects with different severity. For instance, let us assume two runs with the same characterization setup of two different campaigns. After the parsing, the first run finally revealed some CEs, and the second run was classified as SDC. To quantify the criticality of the effects of different experimental runs of different campaigns with the same setup, we define the "severity function" $S_v$, where v is the voltage value, as presented below:

$$S_v = W_{SDC} \cdot \frac{SDC}{N} + W_{CE} \cdot \frac{CE}{N} + W_{UE} \cdot \frac{UE}{N} + W_{AC} \cdot \frac{AC}{N} + W_{SC} \cdot \frac{SC}{N}$$

In this function, the parameters *SDC*, *CE*, *UE*, *AC* and *SC* can take the values from 0 to *N* (*N* is the number of runs at each voltage level), and represent the times that this effect appears to these runs. Parameters $W_{SDC}$, $W_{CE}$, $W_{UE}$, $W_{AC}$ and $W_{SC}$ represent "weights" that can be set to characterize the severity of each effect of TABLE II. The higher the weight, the more critical the effect is considered by our function. For the purpose of this paper, we defined the values presented in TABLE III as values for our severity function (any values for the weights can be used).

TABLE III: WEIGHTS USED IN OUR EXPERIMENTS.

| Weight | Value |
|---|---|
| $W_{SC}$ | 16 |
| $W_{AC}$ | 8 |
| $W_{SDC}$ | 4 |
| $W_{UE}$ | 2 |
| $W_{CE}$ | 1 |
| $W_{NO}$ | 0 |

At the end of the parsing step, all the collected results concerning the characterization (according to TABLE II) and the severity function of each run are reported in CSV files.

## V. EXPERIMENTAL EVALUATION

We present indicative examples of the results that can be generated by the characterization framework. As the aim of this paper is to describe and present the automated framework for system-level characterization the results are presented to demonstrate the framework capabilities.

The framework can reveal for each core of the CPU and the evaluated program the three different regions of operation when we reduce the voltage. These are the *safe* and *unsafe* operating regions and the region in which the system cannot operate (*crash*). For our experiments, we used two different benchmarks: *Linpack*, which is a widely-used high-performance benchmark [25] and *hmmer* from the SPEC CPU2006 benchmark suite [26] with the reference input dataset. Both of them ran on a single core in each PMD, while the remaining six cores (that is 3 PMDs) are reliable (see explanation in section IV). In order to present the non-deterministic behavior of such experiments, we ran each campaign three times. Fig. 3 and Fig. 4 present the three campaigns for the Linpack and hmmer benchmark (1st, 2nd, 3rd), respectively, in the case that are executed in each individual core running at 2.4GHz, while the rest of the PMDs operate on the reliable cores setup. In both benchmarks, we can notice the three regions of operation according to the collected results. The regions are:

- *Safe* region (green): The characterization runs that correspond to this zone had a NO (normal operation) without SDCs, errors or crashes.
- *Unsafe* region (yellow): The characterization runs that correspond to this zone manifested an abnormal behavior (SDC, CE, UE, AC) but not a system crash.
- *Crash* region (grey): This region is created by the first characterization runs that lead to a system crash.

We observe significant variation among the three runs, and also significant core-to-core static variation when they execute the same benchmark. From these results, we observe that cores 4 and 5 of the particular chip we used are more robust than the others. Moreover, it is important to notice the width of the *Safe* region in the two benchmarks that is up to 11.2% lower than the nominal voltage value (980mV). This reduction of the voltage from the nominal value corresponds to power gains up to more than 21%. Finally, the width of the *Safe* region ranges from 0mV up to 40mV. This illustrates that with the development of appropriate mitigation techniques the power gain can reach the 28.3%.

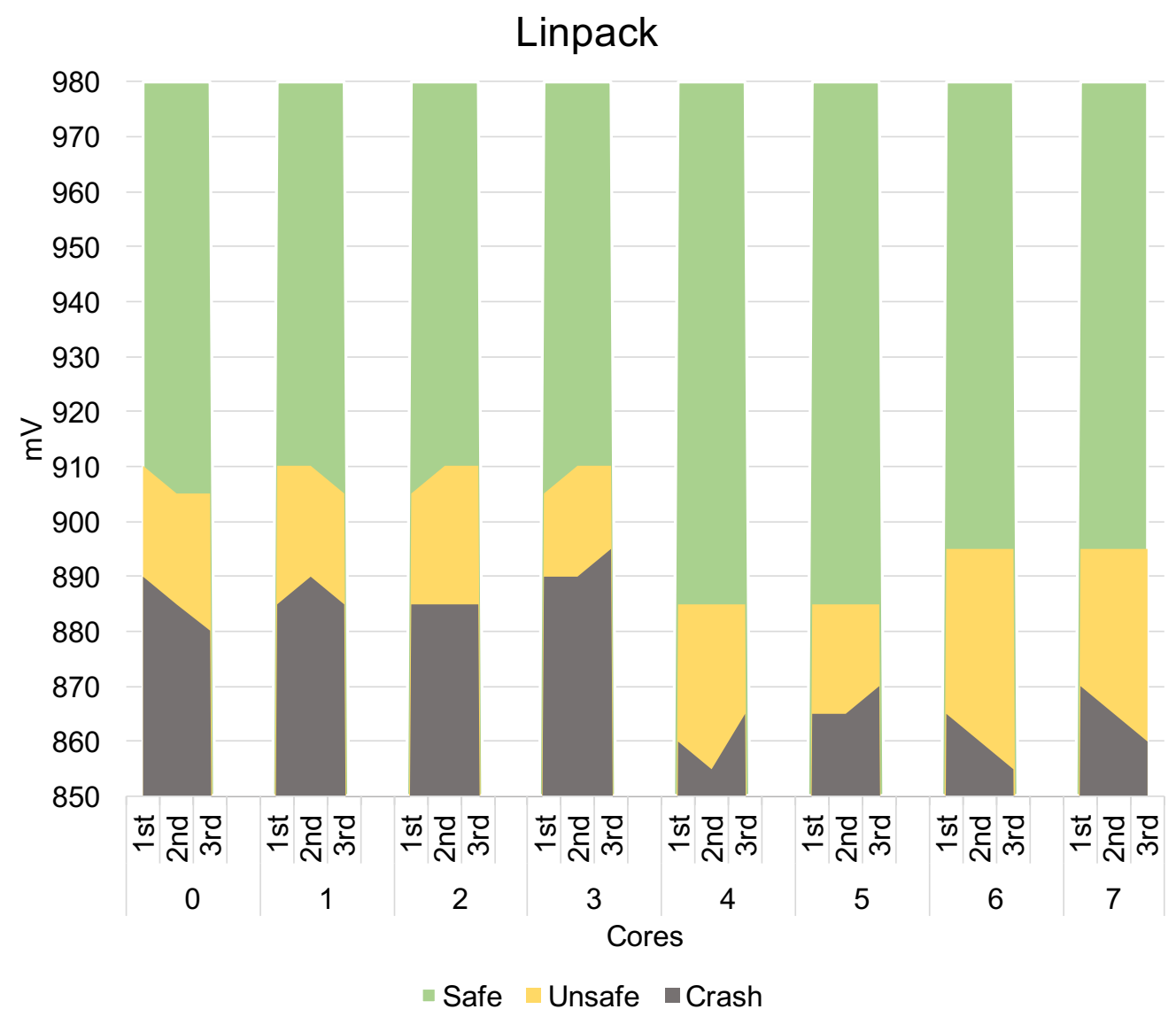

Fig. 3. Linpack benchmark - Cores characterization.

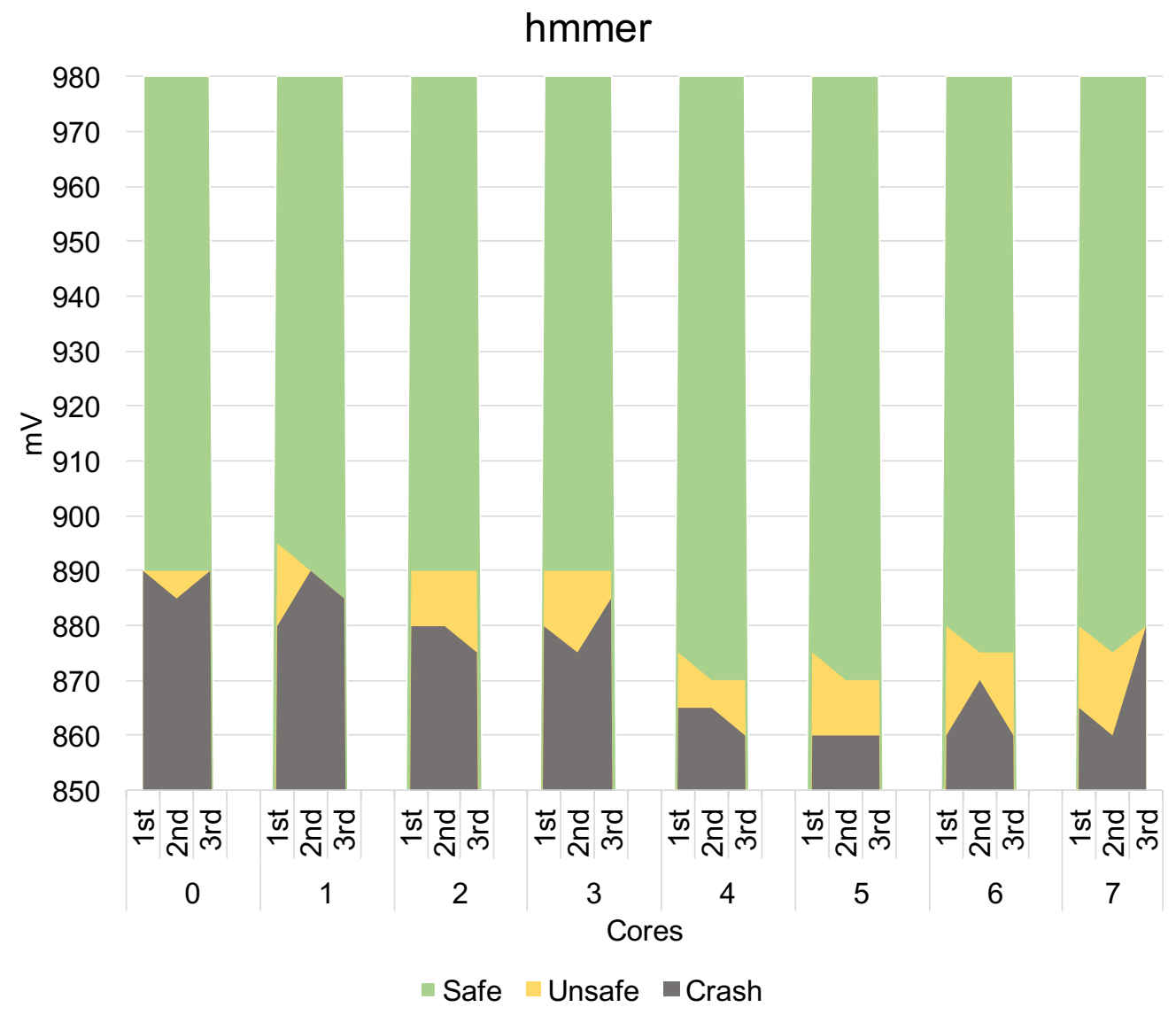

Fig. 4. hmmer benchmark - Cores characterization.

Finally, for the same characterization runs we used the severity function presented in subsection IV.C to present the scaling of the effects and their severity in the reduced voltage margins. In Fig. 5 and Fig. 6 we can notice that the lighter the color, the more stable and reliable is the system. While reducing the voltage margins, we observe that the instability increases (the color becomes darker), until the darkest color, which indicates that the system cannot operate in such voltage margins.

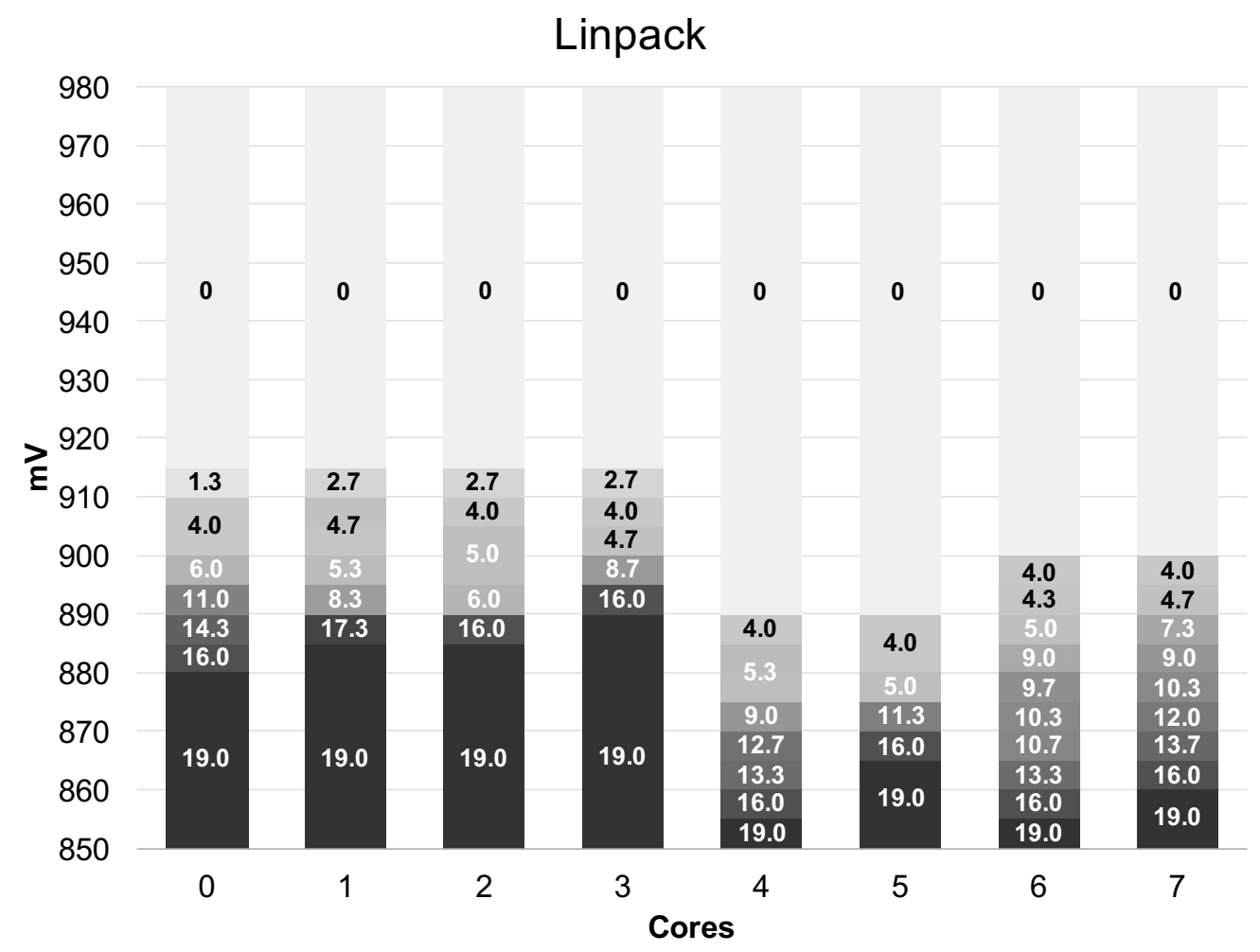

Fig. 5. Linpack benchmark - Severity scaling.

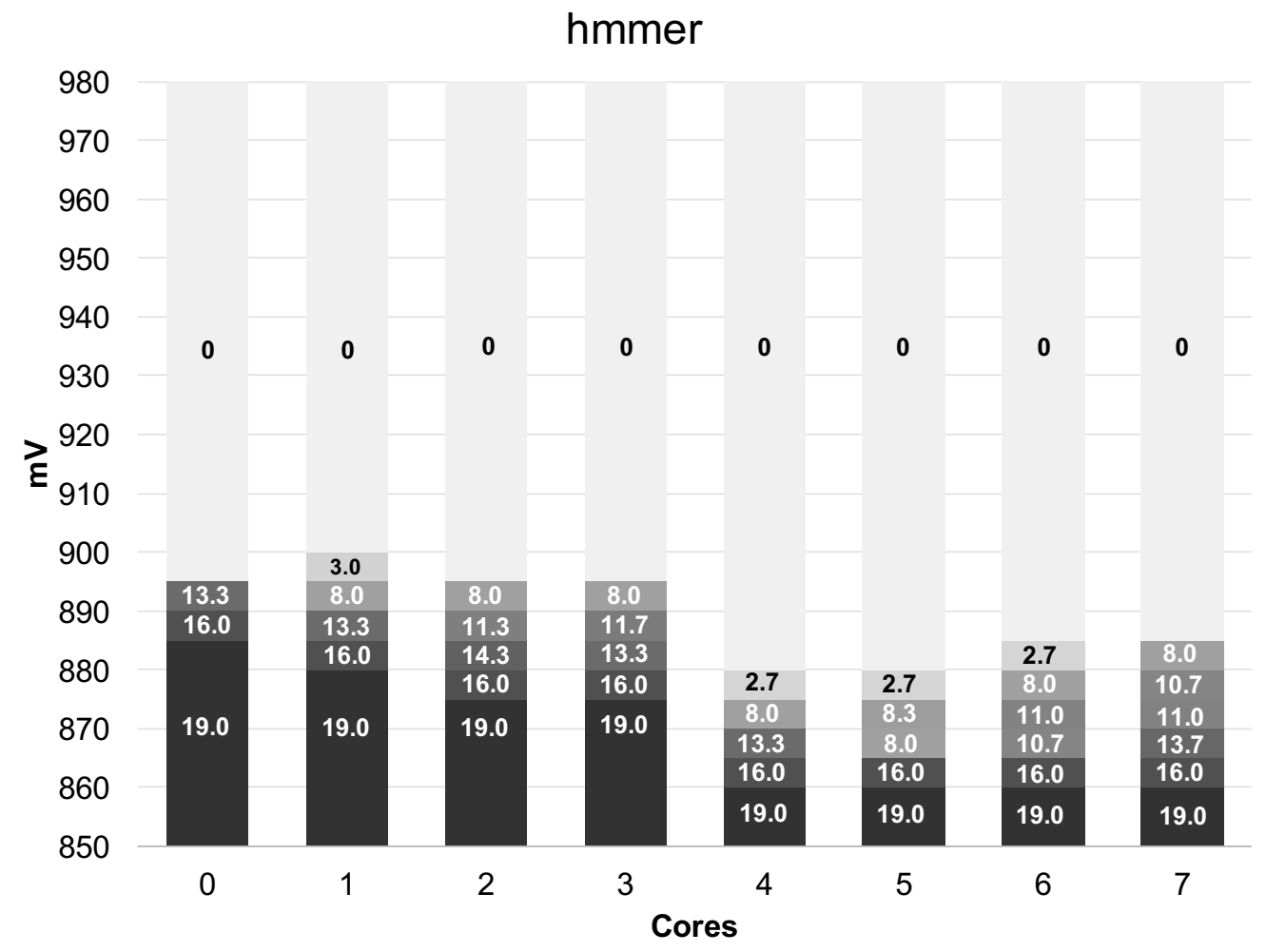

Fig. 6. hmmer benchmark - Severity scaling.

## VI. CONCLUSION

We presented a versatile framework for system-level voltage and frequency scaling characterization built on top of the ARMv8-compliant APM's X-Gene 2 micro-server family. The framework is fully automated and reports information supported by the hardware itself, such as cache ECC errors, as well as SDCs, system or process crashes, and hangs. We present the challenges for the development of such a framework, and describe how we overcame these challenges, by using and combining several hardware, software and system engineering techniques. We also proposed a new metric for both aggregating the results produced by multiple runs due to non-deterministic executions, and quantifying the microprocessor's ability to operate beyond nominal conditions. Finally, we presented some experimental results, which demonstrate potential uses of the characterization framework to identify the limits of operation and support system-level design decisions for improved energy efficiency.

## Acknowledgment

This work is funded by the H2020 Framework Program of the European Union through the UniServer Project (Grant Agreement 688540).